\documentclass[aps, prd, amsmath, floats, floatfix, onecolumn, superscriptaddress, nofootinbib, 10pt]{revtex4-2}
\usepackage{eufrak}
\usepackage{dsfont}
\usepackage{amssymb}%relation symbols

\usepackage{physics}
\newcommand{\be}{\begin{eqnarray}}
\newcommand{\ee}{\end{eqnarray}}
\usepackage{slashed} %Feynman Slash Notation

\usepackage{color}
\definecolor{ao(english)}{rgb}{0.0, 0.5, 0.0}
\usepackage[breaklinks,pdfpagelabels]{hyperref} %''hypertex'' option to compile in dvi without errors
\hypersetup{colorlinks=true,linkcolor=blue,urlcolor=blue,citecolor=ao(english)}

\usepackage{cleveref}
\crefname{equation}{}{}

\usepackage{booktabs}
\usepackage{multirow}
\usepackage{threeparttable}

\usepackage[]{mdframed}

\usepackage{xcolor}

\usepackage{graphicx}

\begin{document}

\title{Acausal exact vacuum solutions in nonlocal gravity}

\author{Zhe~Zhao}
\email[]{zzhao24@m.fudan.edu.cn}
\affiliation{Center for Astronomy and Astrophysics, Department of Physics, Fudan University, Shanghai 200438, China}

\author{Leonardo~Modesto}
\email[]{leonardo.modesto@unica.it}
\affiliation{Dipartimento di Fisica, Universit\`a di Cagliari, Cittadella Universitaria, 09042 Monserrato, Italy}
\affiliation{I.N.F.N, Sezione di Cagliari, Cittadella Universitaria, 09042 Monserrato, Italy}

\author{Cosimo~Bambi}
\email[Corresponding author: ]{bambi@fudan.edu.cn}
\affiliation{Center for Astronomy and Astrophysics, Department of Physics, Fudan University, Shanghai 200438, China}
\affiliation{School of Natural Sciences and Humanities, New Uzbekistan University, Tashkent 100000, Uzbekistan}

\date{\today}

\begin{abstract}
Nonlocal gravity is a promising super-renormalizable or finite quantum gravity theory consistent with unitarity. In this paper, we focus on the classical equations of motion and explicitly show that a particular subclass of G\"{o}del-type Universes, where closed time-like curves are allowed, is an exact solution of nonlocal gravity in vacuum. The result is consistent with a well defined theory at quantum level, but it is realized only with a special, although large, class of nonlocal form factors. Therefore, by itself the renormalizability requirement is not a sufficient guiding principle in vacuum whether we want to avoid the causality violation. From the physical point of view, the causality violation takes place from the non locality fundamental scale to macroscopic scales. Therefore, it is the presence of matter to break the classical degeneracy between the Minkowski and the G\"{o}del Universe. Finally, we have shown that at the non-perturbative quantum level the transition from a flat to a G\"{o}del Universe is ridiculously small.
\end{abstract}

\maketitle
%\tableofcontents

\section{Introduction}

Motivated by the pioneering work of Nascimento et al.~\cite{Nascimento:2021bzb} and a previous paper \cite{Zhao:2023tox}, we focus on finding G\"odel-type exact solutions~\cite{Godel:1949ga, Raychaudhuri:1980fd, Reboucas:1982hn, Teixeira:1985wd, Istvan} in a quite general class of nonlocal gravitational theories.

%motivation of NLG
Einstein's gravity based on General Relativity has achieved remarkable success on macroscopic scales. However, at quantum level it has plagued by an infinite number of divergences that makes it non-renormalizable, namely in order to remove all the divergences we have to introduce an infinite number of new operators.

To address this issue, a natural approach in the quantum field theory framework consists in extending the Einstein-Hilbert action principle introducing other local \cite{Avramidi:1985ki,Buchbinder:1992rb,shapiro3,Modesto:2015ozb, Modesto:2016ofr,Liu:2022gun, Rachwal:2021bgb, Anselmi:2025uzj, Anselmi:2017ygm} 
or nonlocal operators \cite{Krasnikov, kuzmin, modesto, Briscese:2024tvc, Modesto:2013oma, Modesto:2012ys, review, modestoLeslaw,Calcagni:2017sov,Calcagni:2018pro, StabilityMinkAO, StabilityRicciAO}. 
%%%
%%%
These theories are super-renormalizable or finite consistently with unitarity \cite{Lee:1969fy,Lee:1970iw,Anselmi:2017yux,Anselmi:2017lia,Anselmi:2018kgz,Anselmi:2021hab,
Anselmi:2022qor, Briscese:2018oyx,Briscese:2021mob}, while macrocausality in the Regge eikonal limit was proved in \cite{Giaccari:2018nzr}.

%history of NLG
The minimal nonlocal gravitational theory was initially proposed by 
Krasnikov~\cite{Krasnikov} in $1987$ on the base of tree-level unitarity, namely a precise choice of non locality was introduced in order to avoid ghost-like degrees of freedom.  
In its original proposal, the asymptotically exponential form factor in the Krasnikov model was not compatible with the power-counting renormalizability theorems. Therefore, two years later Kuz'min~\cite{kuzmin} proposed the same theory but including an asymptotically polynomial form factor. 
Such crucial improvement made the theory surely super-renormalizable and tree-level unitary. 
Only $22$ years later, the theory was noticed by Modesto in the intent of looking for gravitational models for regular black holes 
\cite{modesto, Modesto:2010uh, Bambi:2016wmo, Burzilla:2020bkx, Burzilla:2020utr, Zhou:2023lwc, Burzilla:2023xdd, dePaulaNetto:2023cjw, Zhou:2022yio, Mo:2022szw, dePaulaNetto:2021axj,Giacchini:2021pmr, Boos:2021suz, Boos:2020qgg}. Afterwards, in order to get a finite theory of quantum gravity Modesto extended the original minimal theory to any dimension \cite{Modesto:2012ys}, thus showing to be free of ultraviolet divergences in odd dimension \cite{Modesto:2013oma}. Finiteness in any dimension was achieved in collaboration with Rachwal in \cite{modestoLeslaw}. 
Nonlocal gravity (NLG) was generalized to all fundamental interactions in \cite{Modesto:2021ief, Modesto:2021okr} and the quantum properties studied in 
\cite{Calcagni:2023goc}.

However, it was later realized that the problem of singularities is not strictly related to the operators that define the theory but to its symmetries. The symmetry that tames singularities is Weyl conformal symmetry \cite{SC,Modesto:2019cvh, Zhou:2019hqk, Zhang:2018qdk, Bambi:2017ott, Bambi:2017yoz, Modesto:2017uji, Bambi:2016yne}, and quantum gravity comes into play only in preserving it \cite{modestoLeslaw}.

%history of godel-type metric
%In a parallel vein, 
In order to look for observational implications of the theory,
the search for exact solutions of the field equations of motion (EoM) remains a central pursuit in classical NLG. % for probing the deep implications of a theory.
For the case of Einstein's gravity, 
the G\"odel Universe~\cite{Godel:1949ga}, an exact solution to Einstein's field equations with a negative cosmological constant and a dust source, is known especially for admitting closed time-like curves (CTCs), thus presenting a profound challenge to our understanding of the Universe causal structure.
This has motivated the study of a broader class of G\"odel-type metrics. The conditions for these metrics to exhibit spacetime homogeneity (that is, to be homogeneous in time and space, actually referred to as ST-homogeneous) were systematically established by Raychaudhuri et al.~\cite{Raychaudhuri:1980fd} and Rebou\c{c}as et al.~\cite{Reboucas:1982hn}. Such metrics are characterized by two parameters: $m$ and $\omega$, and their full set of isometries was explicitly classified by Teixeira et al. in \cite{Teixeira:1985wd}.

%ST-homogeneous

In a very interesting and stimulating paper, the causal G\"odel-type metrics in NLG were first considered by Nascimento et al.~\cite{Nascimento:2021bzb}. Afterwards, in the footsteps of \cite{Nascimento:2021bzb}, Zhao et al.~\cite{Zhao:2023tox} 
showed that if a G\"odel-type metric allowing for CTCs is an exact solution of a NLG, then, such theory violates the conditions for been renormalizable at quantum level. In oder words, in \cite{Zhao:2023tox} the authors proved a kind of inconsistency between renormalizability and causality violation.

%our work: godel-type metric is a vacuum solution of NLG
In order to check whether the renormalizability condition could be sufficient to exclude the presence of acausal exact solutions, in this paper we consider the EoM for a class of nonlocal gravitational theories in vacuum, namely in absence of any kind of fluid or cosmological constant.

By replacing the general G\"odel-type metric ansatz into the EoM in a quite general class of nonlocal gravitational theories and simplifying them making use of the constraints coming from renormalizability, we prove that the metrics identified by the paramener $m=0$ are exact vacuum solutions. 
 This result is twofolds: first, it provides a new non-trivial exact solution of the theory, second, since the $m=0$ solution allows for CTCs, the renormalizability condition in vacuum (pure gravity without matter) is not enough to ensure causality at the level of the action principle.

%structure
In section~\ref{sec.godel.metric}, we introduction to the G\"{o}del-type metrics: classification and causal structure. 
In section~\ref{sec.nlg} we recap a class of nonlocal gravitational theory in two different bases: the Ricci-Weyl and the Ricci-Riemann bases, along with the theoretical constraints imposed by renormalizability. In section~\ref{sec.solve}, we replace the  G\"{o}del-type metric into the vacuum equations of NLG, and we obtain the corresponding vacuum solution. In section~\ref{sec.CTCatmacroscopic} we discuss physical bounds about classical and quantum macrocausality violation. Finally, we summarize the results in section~\ref{sec.conclusion}.

%notation
In this paper, we adopt the unit system $c = 1$ and use the metric signature $(-,+,+,+)$.

\section{G\"{o}del-type metrics}
\label{sec.godel.metric}

In $1949$, G\"{o}del~\cite{Godel:1949ga} obtained an ST-homogeneous cosmological solution with CTCs by solving Einstein's field equations in presence of a negative cosmological constant and a dust source. The metric can be written in the following form\footnote{The function \(F\) plays the role of the commonly used G\"{o}del-type metric function \(H\). We use \(F\) instead of \(H\) to avoid the reader getting confued with the function {\rm H} that will appear later when we will introduce the form factors for NLG.},
\be
\label{eq.godeluniverse}
\dd{s}^2=-\qty[\dd{t}+F(x)\dd{y}]^2+D^2(x)\dd{y}^2+\dd{x}^2+\dd{z}^2 \, , \quad 
F(x)=e^{mx},\quad  D(x)=\frac{e^{mx}}{\sqrt2} \, , 
%\label{Goedel}
\ee
and it is sourced by the following energy-momentum tensor, 
\be\label{eq.godel.old}
T_{ab}=\rho V_aV_b, \quad V^a= \qty(\pdv{t})^a,\quad m^2=-2\Lambda_{\rm cc}=\kappa^2\rho=2\omega^2 \, , 
\ee
where $\rho$ is a constant energy density, $V^a$ is the four-velocity of matter, $\Lambda_{\rm cc}$ is the negative cosmological constant, $\kappa^2=8\pi G$, $\omega$ is the rotation speed of matter. Afterwards, Ozsv\'ath~\cite{Istvan} proved the G\"{o}del theorem using spinor techniques, which states that the ST-homogeneous solutions of the Einstein field equations in presence of an ideal fluid are either the G\"{o}del Universe ($\omega\neq0$) or the Einstein static universe ($\omega=0$).

An equivalent form of the G\"{o}del metric \eqref{eq.godeluniverse} in cylindrical coordinates reads: 
\be 
\label{type_godel} 
\dd{s^2}=-[\dd{t}+F(r)\dd\theta]^2+D^{2}(r)\dd\theta^{2}+\dd{r}^2+\dd{z}^2\,.
\ee
In the new coordinates the cylindrical symmetry of the metric \eqref{type_godel} is evident, but its ST-homogeneity is no longer manifest.

Raychaudhuri et al. in~\cite{Raychaudhuri:1980fd} provided the sufficient conditions for the metric in \eqref{type_godel} in order to be ST-homogeneous, namely: 
\be 
\label{m-omega} 
\frac{F'(r)}{D(r)}=2\omega\in\mathbb{R}\backslash\{0\}\qand\frac{D''(r)}{D(r)}=m^2\in\mathbb{R}\,, \label{ST}
\ee
where the prime stays for the derivative with respect to the radial coordinate $r$.
Later, calculations by Rebou\c{c}as et al.~\cite{Reboucas:1982hn} showed that the sufficient condition by Raychaudhuri et al. is also a necessary condition.
Since then, the metric of the form in \eqref{type_godel} that is ST-homogeneous is referred to as a G\"{o}del-type metric.
However, the Killing vector fields provided by Rebou\c{c}as et al. did not constitute a complete set. 
Such issue was later figured out by  Teixeira~\cite{Teixeira:1985wd}.

The cylindrical symmetry and homogeneity, namely the metric (\ref{type_godel}) together with homogeneity, imply the conditions (\ref{m-omega}) on the functions present in the metric (\ref{type_godel}). 
Hence, by solving the differential equations \eqref{ST} we can collect the G\"{o}del-type solutions into four classes, which are explicitly represented by the parameters $m^2$ and $\omega$, namely 
\begin{enumerate}
\item the \textit{hyperbolic class}: $m^2>0$, $\omega\neq 0$:
\begin{eqnarray}
F(r)=\frac{2\omega}{m^2}[\cosh(m \, r)-1]\qand D(r)=\frac{1}{m}\sinh(m \, r),
\end{eqnarray}
\item the \textit{trigonometric class}: $-\mu^2=m^2<0$, $\omega\neq 0$:
\begin{eqnarray}
F(r)=\frac{2\omega}{\mu^2}[1-\cos(\mu \, r)]\qand D(r)=\frac{1}{\mu}\sin(\mu \, r),
\end{eqnarray}
\item the \textit{linear class}: $m^2=0$, $\omega\neq 0$:
\begin{eqnarray}
F(r)=\omega \, r^2\qand D(r)=r \, , 
\label{linear}
\end{eqnarray}
\item the \textit{degenerate class}: $m^2\neq 0$, $\omega=0$:
\begin{eqnarray}
F(r)=0 \, .
\label{degenerate}
\end{eqnarray}
\end{enumerate}
In particular, the G\"{o}del Universe in (\ref{eq.godeluniverse}) corresponds to case $m^2 = 2\omega^2 > 0$.

Since the spacetime metric is completely determined by the parameters $m^2$ and $\omega$, its causal structure is also fully characterized by these two parameters. Specifically, the metric allows for CTCs accroding to:
\be
\label{CTC}
\mathrm{CTC} \quad \Longleftrightarrow \quad 4\omega^2-m^2>0\,.
\ee
The G\"{o}del-type solution of NLG that we will obtain in this paper will lie to the $3$rd class, namely $m^2 = 0$, $\omega \neq 0$. Therefore, according to (\ref{CTC}) the spacetime will allow for the existence of CTCs. For future reference, we here explicit write the line element,
\be 
\label{type_godelCTC} 
\dd{s^2}= - (\dd{t}+ \omega \, r^2 \dd\theta)^2+r^2\dd\theta^{2}+\dd{r}^2+\dd{z}^2\,.
\ee

\section{Nonlocal gravity}
\label{sec.nlg}

%action

In this paper we look for exact vacuum solution of the following NLG in the Ricci-Weyl basis, 
\be 
\label{action} 
S = \int \dd x^4\sqrt{-g}\qty[\frac{1}{2\kappa^2}R+R\gamma_{0}(\square_{\Lambda_*} )R+R_{ab}\gamma_{2}(\square_{\Lambda_*})R^{ab}+C_{abcd}\gamma_{4}(\square_{\Lambda_*})C^{abcd}]\,, 
\ee
where the analytic form factors $\gamma_i$ can be expressed as infinite power series of the dimensionless d'Alembert operator $\square_{\Lambda_*}\equiv \square/\Lambda_*{}^{2}$, i.e.,
\begin{equation}
\gamma_{i}(\square)=\sum_{n=0}^{\infty}\gamma_{i,n}\square^{n}_{\Lambda_*}\qand i=\{0,2,4\},
\label{fr}
\end{equation}
$\gamma_{i,n}$ are the dimensionless coefficients of the power series in $\square_{\Lambda_*}$, ${\Lambda_*}$ is an invariant fundamental mass scale of NLG, namely the nonlocality scale.

However, NLG has been mainly studied in other basis rather then in the form (\ref{action}), and the form factors determined according to tree-level unitarity (no ghosts condition), and  super-renormalizability \cite{Krasnikov,kuzmin,modesto, Giaccari:2015vfh, Modesto:2017uji, Koshelev:2016xqb}.
In particular, the theory in the in the Ricci-Riemann basis, and the form factors take the form:
\be
&& \mathcal{L} = \frac{1}{2\kappa^2} R  +R\tilde\gamma_{0}(\square)R+R_{ab}\tilde\gamma_{2}(\square)R^{ab}+R_{abcd}\tilde\gamma_{4}(\square)R^{abcd}  \, , 
\label{tildeF}\\
&& \nonumber \\
&& \tilde\gamma_0(\Box) =-\frac{(D-2)\qty(e^{\mathrm H_0(\Box)}-1)+D\qty(e^{\mathrm H_2(\Box)}-1)}{  4(D-1)\Box}+ \tilde\gamma_4(\Box) \, , \label{gamma0} \\
&& \nonumber \\
&& \tilde\gamma_2(  \Box)=\frac{e^{\mathrm H_2(\Box)}-1}{ \Box}-4\tilde\gamma_4(\Box) \, . 
\label{gamma2p}
%\label{BasesRiem} 2\kappa^2 \,
\ee
The relations between the two different bases can be found in the Appendix \ref{RB}, i.e. 
\begin{equation}
	\label{rienbasis2}
	\gamma_2+2\gamma_4 =\tilde\gamma_2+4\tilde\gamma_4  \, ,
\end{equation}
which guarantees the super-renormalizability of the theory.

In order to avoid infrared nonlocality, the functions $\mathrm{H}_0$ and $\mathrm{H}_2$ in \eqref{gamma0} and \eqref{gamma2p}, are non-zero entire functions that satisfy $\mathrm{H}_0(0) = \mathrm{H}_2(0) = 0$. 
For the sake of completeness, we here remind the Kuzmin's form factor ${\rm H}(z)$ 
\cite{kuzmin,modesto}:
\be
{\rm H}(z)  =  \int_0^{p(z) } \dd w \, \frac{1 - e^{- w}}{w}
=  \gamma_E + \Gamma[ 0, p(z) ] + \log [ p(z) ] \, , %\quad {\rm Re} \, p(z) > 0 \, , \, , 
\label{Hk}
% see matehmatica file: Kuzmin_Form_Factor.nb
\ee
where $p(z)$ is the most general polynomial of degree $n+1$ in the variable $z$, namely 
\be
p(z) = a_0 + a_1 z + a_2 z + \dots + a_{n+1} z^{n+1} \, , \,\,\, a_i \in \mathbb{R} \, . 
\label{Poly}
\ee
Later, we will provide explicit examples of ${\rm H}_i$ ($i=1,2$) and $p(z)$.

In addition to the condition (\ref{rienbasis2}), we will consider form factors $\gamma_i$ such that: 
\be
\label{condition1} \gamma_2(0)+2\gamma_4(0)=0 \, .
\ee
Therefore, according to (\ref{rienbasis2}) in the Ricci-Riemann basis we get:
\be
\label{condition2} 
\tilde\gamma_2(0)+4\tilde\gamma_4(0)=0 \quad \overset{\eqref{gamma2p}}{ \Longrightarrow} \quad \eval{ \frac{e^{\mathrm H_2(x)}-1}{x}}_{x=0}=0
\quad 
\Longrightarrow 
\quad 
\mathrm H_2(0)=0 \, , \quad 
\mathrm {\rm H}'_2(0)=0 \,. 
\ee

\section{Vacuum solution}

\label{sec.solve}

In this section, we are going to show that the G\"{o}del-type metric with $m^2 = 0$ is an exact solution of the nonlocal theory (\ref{action}) consistently with the conditions (\ref{condition1}) and (\ref{condition2}).

For future reference and to simplify the equations we introduce the following function,
\be
\label{ff-exp} \label{ffunction} 
f(x)=\gamma_2(x)+2\gamma_4(x)=\tilde\gamma_2(x)+4\tilde\gamma_4(x)=\frac{e^{\mathrm H_2(x)}-1}{x}\,. 
\ee
Therefore, the conditions  (\ref{condition1}) and (\ref{condition2}) simplify to:
\be
\label{condition3} 
f(0)=0\qand {\rm H}'_2(0)=0\,.
\ee

\subsection{EoM of NLG}
\label{sec.eom.nlg}
The EoM for the theory (\ref{action}) were derived in~\cite{Zhao:2023tox},
but can be found also in the Appendix \ref{RB2} of this paper too. 
In short notation, the EoM read:
\be  G_{ab} +Q_{ab}=0 \,.
\label{EoM2}
\ee
where $Q_{ab}$ arises from the variation of the last three terms of the action \eqref{action}. 
Additionally, we adopt the convention that Greek letters represent components of tensors, and we use the following orthonormal basis:
\be 
\label{basis} 
g_{ab}=\eta_{\mu\nu}(e^\mu)_a(e^\nu)_b\qand  \{(e^{\mu})_a\} = \{\dd{t}+F(r)\dd{\theta}, \dd r,D(r)\dd\theta,\dd z  \} \,. 
\ee

In order to show that the G\"{o}del-type metric can be an axact solution of the theory (\ref{action}), we replace the specific form of the metric into the EoM. Finally, the EoM can then be written in the following very simple form~\cite{Zhao:2023tox}, 
\be\label{EoM}
G_{\mu\nu} +Q_{\mu\nu}=0\,,
\ee
where the Einstein's tensor reads:
\be
\label{Einsteintensor} 
 G_{\mu\nu}=\mqty(\dmat{3\omega^2-m^2,\omega^2,\omega^2,m^2-\omega^2}) \,,
\ee
while the $Q_{\mu\nu}$ tensor is: 
\be
&&	\hspace{-1.1cm} 
\label{QINF}
Q_{\mu\nu}=\nonumber2\kappa^2\frac{4\omega^2}{{\Lambda_*}^2}(m^2-4\omega^2)^2f'\qty(\frac{6\omega^2}{{\Lambda_*}^2})\mqty(\dmat{1,1,1,0}) \\
&& \hspace{1.7cm}
+2\kappa^2\frac{m^2-4\omega^2}{3}f\qty(\frac{6\omega^2}{{\Lambda_*}^2})\mqty(\dmat{m^2-20\omega^2,m^2-12\omega^2,m^2-12\omega^2,4\omega^2-m^2}) \, , 
\ee
where $f(x)$ is defined in \eqref{ffunction}.
Combining \eqref{Einsteintensor} and \eqref{QINF}, the EoM \eqref{EoM} simplify to:
\be
\label{1}
3\omega^2-m^2 +2\kappa^2\qty[\frac{4\omega^2}{{\Lambda_*}^2}(m^2-4\omega^2)^2f'(6\omega^2/{\Lambda_*}^2)+\frac{m^2-4\omega^2}{3}f(6\omega^2/{\Lambda_*}^2)(m^2-20\omega^2)]= 0\, , \\
\label{2}
\omega^2 +2\kappa^2\qty[\frac{4\omega^2}{{\Lambda_*}^2}(m^2-4\omega^2)^2f'(6\omega^2/{\Lambda_*}^2)+\frac{m^2-4\omega^2}{3}f(6\omega^2/{\Lambda_*}^2)(m^2-12\omega^2)]=0 \, , \\ 
\label{3}
m^2-\omega^2  -2\kappa^2\qty[\frac{(m^2-4\omega^2)^2}{3}f(6\omega^2/{\Lambda_*}^2)]= 0 \, .
\ee
Introducing the following definitions,  
\be z \equiv \frac{ m^2-4\omega^2}{{\Lambda_*}^2} \, \quad x\equiv \frac{\omega^2}{{\Lambda_*}^2} \, ,
\label{zxmo}
\ee
 and the dimensionless parameter $\lambda\equiv 1/\qty(2\kappa^2{\Lambda_*}^2)$, we can simplify further the EoM \eqref{1}, \eqref{2}, and \eqref{3} as follows:
\be
\label{1'}
(\ref{1})-(\ref{2})&\Longrightarrow& \qty[\lambda+\frac{8}{3}xf(6x)]z+2\lambda x =0\,,\\
\label{2'}
(\ref{2})&\Longrightarrow& \qty[\frac{f(6x)}{3}+4xf'(6x)]z^2-\frac{8}{3}xf(6x)z+\lambda x=0\,,\\
\label{3'}
(\ref{3})&\Longrightarrow& -\frac{f(6x)}{3}z^2+\lambda z+3\lambda x=0\,.
\ee
The relationship between the dimensionalities of the physical quantities mentioned above is as follows:
\be
\label{dimension}
\qty[\frac{1}{\kappa^2}] =\qty[{\Lambda_*}^2]=\qty[m^2]=\qty[\omega^2]\,.
\ee

So far we have three equations \eqref{1'}, \eqref{2'}, and \eqref{3'}, for two variables $x$ and $z$, and in turn for $m$ 
and $\omega$. We first solve equation \eqref{1'} for $z$ after defining  
\be
k=4f(6x)x\, ,
\ee 
the result is:
\be
\label{1''}
\eqref{1'} \quad \Longrightarrow \quad z=-\frac{2\lambda x}{\lambda+2k/3 }\,.
\ee
Now, by replacing \eqref{1''} into \eqref{3'}, we get the following equation for $k$, 
\be
\label{v-solution} 
\eqref{1''} \,\, \rightarrow \,\, \eqref{3'} \quad \Longrightarrow \quad 3\lambda k+4k^2=0\,.
\ee
Therefore, a nontrivial solution (the other solution would be $k=0$) of  \eqref{v-solution}  
reads:
\be
\label{3''}
k=-\frac{3}{4}\lambda\,=4f(6x)x .
\ee
Replacing \eqref{1''} and \eqref{3''} into  \eqref{2'} and performing simple simplifications, we get:
\be
\label{2''}
\lambda=32x^2f'(6x)\,.
\ee
Therefore, EoM \eqref{1'}, \eqref{2'}, and \eqref{3'} simplify to: 
\be
&& \label{fin1}
z+4x=0 \, , \quad  \\
&& -\frac{3}{4}\lambda=4f(6x)x \, , \label{34} \\
&&  \lambda=32x^2f'(6x) , 
\label{32}
\ee
where to got the first equation by replacing $\lambda = - 4 k/3$ (from (\ref{3''})) into (\ref{1''}).

Now, using $f(x)$ defined in \eqref{ff-exp}, $x$, and $z$, we can further simplify \eqref{fin1},\eqref{34}, and (\eqref{32}), namely:
\be
&& \nonumber 
(\ref{zxmo}) \qand  z+4x=0 \quad \Longrightarrow \quad m^2=0 \, , \\
&& (\ref{34}) \qand \eqref{ff-exp} 
\quad \Longrightarrow \quad  \eval{e^{\rm H_2(\zeta)}}_{\zeta=6x} =1-\frac{9}{8}\lambda \, , \nonumber \\ 
&& 
(\ref{34}), (\ref{32}) \qand \eqref{ff-exp} \quad \Longrightarrow \quad  \eval{{\rm H_2}'(\zeta)}_{\zeta=6x}=0 \, .
\label{fin2}
\ee
Based on the simplified EoM (\ref{fin2}), we observe that whether the G\"{o}del-type metric is a solution of the NLG depends lonely on $\mathrm{H}_2$. In the next subsection, we will focus on possible choices of $\mathrm{H}_2$ and we provide examples of entire function $\mathrm{H}_2$ consistent with the G\"{o}del-type metric for $m^2=0$, namely consistent with the conditions (\ref{fin2}) and (\ref{condition3}). 

Let us recap the conditions need the G\"{o}del-type metric to be an exact solution, namely condition (\ref{fin2}) and (\ref{condition3})
\be
&& \boxed{
f(x)= \frac{e^{\mathrm H_2(x)}-1}{x}\, , \quad 
f(0)=0\qand {\rm H}'_2(0) = 0 
\, ,  
}
\label{prima} \\
&&
\boxed{
m^2=0 \, , \quad \eval{e^{\rm H_2(\zeta)}}_{\zeta=6x} =1-\frac{9}{8}\lambda \, , \quad 
 \eval{{\rm H_2}'(\zeta)}_{\zeta=6x}=0 \quad \mbox{where} \quad x\equiv \frac{\omega^2}{{\Lambda_*}^2} \quad 
 \mbox{and} \quad 
 \lambda \equiv \frac{1}{2 \kappa^2 {\Lambda_*}^2} 
  \,  .
 }
 \label{seconde}
\ee
Given the dimensionless constant $\zeta$ we have:
\be
\omega^2 = \frac{ \zeta{\Lambda_*}^2 }{6} . 
\label{omegaL}
\ee

%Form factors of nonlocal gravity
\subsection{A general class of entire functions ${\rm H}(x)$}
\label{sec.Hx}

A general class of entire functions compatible with unitarity, but in particular with the power counting super-renormalizability \cite{kuzmin,modesto,Briscese:2024tvc}
can be defined through the following integral, 
\be
\label{Hpoly}
{\rm H}(x)\equiv \alpha \int_0^{p(x)}\dd{z}\frac{1-g(z)}{z}\,, 
\ee 
where $p(x)$ is a generic polynomial of $x$ and degree $n$, while $\alpha$ is a positive constant.
In order to achieve an asymptotic polynomial behavior of $\exp \mathrm H(x)$ as request from renormalizability \cite{kuzmin,modesto}, $p(x)$ and $g(x)$ in \eqref{Hpoly} must satisfy the following requirements~\cite{kuzmin,modesto,Briscese:2024tvc}: \
\begin{enumerate}
\item $p(x)$ is a real polynomial such that  $p(0) = 0$. 
\item $g(z)$ is an entire and real function on the real axis with $g(0) = 1$,
\item  $|g(z)| \rightarrow 0$ for $|z| \rightarrow + \infty$ in a conical region around the real axes. 
\end{enumerate}

As an example, let us consider the following function $g(z)$, 
\be
\label{eq.gz} g(z)=\exp(-z^n)\, .
\ee   
Therefore, the result of the integral (\ref{Hpoly}) gives the following entire function,
\be
\label{Hx}
{\rm H}(x)=\frac{\alpha}{n}\qty{\ln p^n(x)+\Gamma[0,p^n(x)]+\gamma_{\rm E}}\,,
\quad \mbox{where} \quad 
 \Gamma[0,z]=\int_z^{+\infty}\dd{z'}\frac{e^{-z'}}{z'}\, , 
\ee
and $\gamma_{\rm E}$ is the Euler-Mascheroni constant. 
The limit of (\ref{Hx}) for $p(x)\rightarrow0$ is:
\be
\label{asymtotic}
{\rm H}(x)=\frac{\alpha}{n}p^n(x) \quad \Longrightarrow \quad {\rm H}(0)=0 \, .
\ee

According to the value of the integer $n$, we are going to 
discuss for which entire functions $\mathrm{H}_2$ the G\"{o}del-type metric with $m^2=0$ is an exact solution of NLG.

\subsubsection{The $n=1$ case}
\label{sec.n=1}
%%%
%\begin{itemize}
	 %Kuz'min form factor~\cite{Kuzmin:1989sp}\\
		Let us start considering the Kuz'min form factor~\cite{kuzmin}, namely the case $n = 1$ in (\ref{Hx}), $p(x) = x$, and $\alpha \geq 3$. In this case, $\mathrm{H}_2$ simplifies to:
\be
{\rm H}_2(x)=\alpha\qty{\ln x+\Gamma[0,x]+\gamma_E}\, .
\ee 
Therefore, one can show that: 
\be {\rm H_2}'(0)\sim p'(0)\neq0\,, 
\ee 
which contradicts our assumption \eqref{prima}. Therefore, in order to be consistent with our requirement \eqref{prima}, $p(x)$ must have at least a double root in $x = 0$, for example: $p(x) = x^2$.

Suppose that, through an appropriate choice of $p(x)$, the condition \eqref{prima} has already been satisfied. We now proceed to evaluate the condition under which \eqref{seconde} holds. 
Let us begin evaluating the derivative of ${\rm H_2}$ are request in the third condition in (\ref{seconde}),  \be {\rm H_2}'(\zeta)=\alpha \eval{\frac{1-\exp(-z)}{z}}_{z=p(\zeta)} p'(\zeta)\,. \ee Therefore according to the third equation in (\ref{seconde}), namely ${\rm H_2}'(\zeta)=0$, we get: \be\label{eq.n=1eom3'} p'(\zeta)=0\,. \ee
At this point, to verify whether the G\"{o}del-type metric is an exact solution with $m^2=0$, it suffices to analyze whether \eqref{eq.n=1eom3'} matches the second equation in \eqref{seconde}.
In order to satisfy  the second equation in \eqref{seconde}, we need:
\be \label{n=1H<0} {\rm H}_2(\zeta = 6 \omega^2/{\Lambda_*}^2) < 0  \quad \mbox{which implies:} \quad 
p(\zeta)<0 \, .
\ee

Therefore, the G\"{o}del-type metric is an exact solution with $m^2=0$ if $p(x)$ has at least a double root at $x=0$ and $\exists \, \zeta$ such that $p'(\zeta)=0$ and $p(\zeta)<0$.

To illustrate the above general analysis on the existence of solutions, we now present an explicit example of a vacuum G\"odel-type solution with \( m^2 = 0 \). Consider $p(x) = x^2 \qty(2x - 3\zeta)$.

First, we note that in the limit \( x \to 0 \), $ p(x) \sim -3\zeta x^2$, which satisfies the condition \eqref{prima}. We next demonstrate that, for an appropriate choice of \( \zeta \), the condition \eqref{seconde} can also be fulfilled. A straightforward calculation gives \( p'(\zeta) = 0 \) and \( p(\zeta) = -\zeta^3 < 0 \); therefore, \( x = \zeta \) corresponds to an extremum of \( \mathrm{H}_2 \), with \( \mathrm{H}_2(\zeta) < 0 \).

To satisfy the condition \( e^{\mathrm{H}(\zeta)} = 1 - \frac{9}{8}\lambda \) in \eqref{seconde}, one must in general determine suitable values of the parameters \( \alpha \) and \( \zeta \). Although this relation is not easily solved in closed form, it can be treated analytically under a simplifying assumption. In particular, if \( \zeta \) is sufficiently small (introduced here solely for illustrative purposes and not required in general), we obtain \be e^{{\rm H}(\zeta)}\simeq 1-\alpha\zeta^3+\order{\zeta^6}=1-\frac{9}{8}\lambda\Longrightarrow \zeta=\qty(\frac{9\lambda}{8\alpha})^{1/3} \,. \ee

Therefore, within the NLG theory specified by \( n = 1 \) and $p(x) = x^2 \bigl(2x - 3\zeta\bigr)$, where $\zeta = \left(\frac{9\lambda}{8\alpha}\right)^{1/3}$, there exists a G\"odel-type solution with \( m^2 = 0 \) and \( \omega^2 = \frac{\zeta}{6}\Lambda_*^2 \).

\subsubsection{The $n$-even case}
\label{sec.n.even}
When $n$ is even, it is easy to check that $\mathrm{H}$ is an even function of $x$ such that:
\be
{\rm H}_2(x)=\alpha\int_0^{p(x)}\dd{z}\frac{1-e^{-z^n}}{z}\geq0\,.
\ee 
Therefore, $e^{\rm H_2}\geq1$, which contradicts the second condition in \eqref{seconde} because $\lambda$ is strictly positive. Hence, {\it for $n$ even NLG does not have G\"{o}del-type solutions with $m^2=0$}.

\subsubsection{The $n$-odd case ($n\geq3$)}
\label{sec.n.odd}

We first show that when $n$ is odd and $n\geq3$, it does not contradict \eqref{prima}. %condition3}. 
Based on \eqref{asymtotic}, we have
\be
 \lim_{x\to 0}{\mathrm H_2}'(x)\propto p^{n-1}(x) p'(x)\,.
\ee
Since $n \geq 3$, we have ${\rm H_2}'(0) = 0$
consistently with \eqref{prima}. 
Next, we have to check at the conditions (\ref{seconde}). 
Let us begin evaluating the derivative of ${\rm H_2}$ are request in the third condition in (\ref{seconde}), 
\be
{\rm H_2}'(\zeta)=\alpha \eval{\frac{1-\exp(-z^n)}{z}}_{z=p(\zeta)} p'(\zeta)\,.
\ee
Therefore according to the third equation in (\ref{seconde}), namely ${\rm H_2}'(\zeta)=0$, we get:
\be\label{eq.eom3'}
p(\zeta)=0\qor p'(\zeta)=0\,.
\ee

Finally, to verify whether the G\"{o}del-type metric is an exact solution with $m^2 = 0$, it is sufficient to analyze whether \eqref{eq.eom3'} is compatible with the second equation in \eqref{seconde}. 
According to (\ref{eq.eom3'}), we have to consider two cases. 
\begin{itemize} 
	\item Case $1$, $n$ is odd and $p(\zeta)=0$, thus, according to the first definition in (\ref{asymtotic}), i.e. 
	${\rm H}(x)=\frac{\alpha}{n}p^n(x)$, we get 
	${\rm H}_2(\zeta) = 0$. Hence, 
	the second condition in (\ref{seconde}) requires:
	\be
	%\lim_{x \rightarrow 0} 
	{\rm H}_2(\zeta) = 0 \quad \mbox{and}  \quad \eval{e^{\rm H_2(\zeta)}}_{\zeta=6x} =1-\frac{9}{8}\lambda  \quad \Longrightarrow \quad \lambda = 0. \label{eq.case1}
	\ee
which contradicts the condition for a non-trivial NLG, with $\Lambda < \infty$. Hence, 
in this case the G\"{o}del-type metric is not a solution. 
	\item Case $2$, $n$ is odd and $p'(\zeta)=0$. 
In order to satisfy 
the second equation in \eqref{seconde}, we need:
%the exponent $H_2(\zetta = 6 \omega^2/\Lambda^2$ 
\be
\label{H<0}
{\rm H}_2(\zeta = 6 \omega^2/{\Lambda_*}^2) < 0  \quad \mbox{which implies:} \quad p(\zeta)<0 \, .
\ee

Therefore, according to (\ref{fin2}) and (\ref{H<0}), the G\"{o}del-type metric is an exact solution 
if $\exists \, \zeta$ such that $p'(\zeta)=0$ and $p(\zeta)<0$.

\end{itemize}

\section{CTC at macroscopic scales}
\label{sec.CTCatmacroscopic}
In this section we would like to quantify the violation of causality of a probe particle that does not affect the solution. According to the metric, which we recall here to facilitate its analysis, 
\be 
\label{CTC2} 
\dd{s^2}= - (\dd{t}+ \omega \, r^2 \dd\theta)^2+r^2\dd\theta^{2}+\dd{r}^2+\dd{z}^2\,.
\ee
in order to have a CTC curve the angular coordinate $\theta$ must be time-like. Assuming to be able to travel on a curve with $t$, $r$, and $z$ all constant, the line element turns into:
\be 
\label{CTC2} 
\dd{s^2}= - r^2 (\omega^2 \, r^2 - 1) \dd\theta^2 \, . 
\ee
Hence, according to our signature, when;
\be\label{eq.ctcregion}
\omega^2 \, r^2 - 1 > 0 \ , \quad r > \frac{1}{\omega} \, , 
\ee
the periodic coordinate $\theta$ is time-like and an observer at rest in the space-like coordinates $t,r,z$ follows a CTC. It is useful to stress that the critical radius \eqref{eq.ctcregion} is independent on the nonlocal operators appearing in the theory. Once the linear G\"{o}del-type metric (\(m^2=0\)) is obtained, the onset of CTCs is determined purely by the sign of the metric component $g_{\theta\theta}$. The role of the NLG is instead indirect: the field equations select which values of \(\omega\) are allowed. Using \eqref{omegaL}, $\omega^2=\zeta\Lambda_*{}^2/6$, one may equivalently write:
\be  
r_c=\frac{1}{\omega}=\sqrt{\frac{6}{\zeta}}\ell_p\,, \qquad \ell_p=\Lambda_*^{-1}\,. 
\ee
Thus the nonlocal parameters do not change the causal structure of the spacetime, but they set the physical scale of the critical radius through the admissible values of \(\zeta\) and the nonlocality scale \(\Lambda_*\). In particular, for \(\zeta\sim 1\) and \(\Lambda_*\) of the order of the Planck mass, the CTC region starts at the Planck scale, whereas smaller admissible values of \(\zeta\) push the critical radius to larger distances. Since the CTC region extends over all radii \(r>r_c\), the violation of causality is not merely a local or microscopic effect, rather, it constitutes a macroscopic violation of causality. Let us evaluate the proper time needed an observer to travel through an entire time-like loop. Integrating the angular coordinate from $0$ to $2 \pi$, we get:
\be
\Delta \tau = 2 \pi \,  r \sqrt{\omega^2 r^2 - 1} \, . 
\ee
According to the above formula, if we are at a distance $r = 1$ m from the rotation axis of the Universe the proper time to make a temporal loop is about $\Delta \tau \sim 10^{27}$ sec, while at a distance of one Fermi, namely $r = 10^{-13}$ cm, $\Delta \tau \sim 10^{-3}$ sec. Therefore, subnuclear particles close to the rotation axis of the Universe can experience the time-travel in increasingly shorter times as their Compton length decreases.

A quantum level in the {\em non-boundary proposal}, the wave function of the Universe can be obtained performing the path integral over all possible Euclidean metrics that can be analytically continued to the Lorentzian section \cite{Gibbons:1976ue}. Implementing the saddle point approximation,
\be
\Psi \propto e^{-I} \, ,
\ee
where $I$ is the action evaluate on the instantonic solution. Therefore, the probability for the a Minkowski Universe to decay into the the G\"{o}del-type metric found in this paper is \cite{Carr:2025auw}:
\be
\Gamma = \left| \frac{ e^{-I_{\text{G\"{o}del} }}}{ e^{-I_{\rm Minkowski}}} \right|^2
= e^{-2 I_{\text{G\"{o}del} } + 2 I_{\rm Minkowski}} 
= e^{-2 I_{\text{G\"{o}del} }}
= e^{-2 V_4 M_{\rm p}^4 } = e^{-2 \left(  \frac{L_{\rm Universe}}{ \ell_{\rm p}} \right)^4 } 
\approx e^{- 2 \times 10^{244}} \, ,
\label{ProbGM}
\ee
which is ridiculously small. In (\ref{ProbGM}), we assumed the non locality scale to be of the same order of magnitude of the Planck cale. Moreover, for the radius of the Universe we used $L_{\rm Universe} = 10^{28}$ cm. However, when the radius of the Universe is of Planck size the probability increases until reaching the value of $e^{-2}$.

\section{Conclusions and Summary} \label{sec.conclusion}

In this paper we have extended the previous works~\cite{Nascimento:2021bzb, Zhao:2023tox} to the case of zero energy momentum tensor.
In particular, we have explicitly shown that the G\"{o}del-type metric with $m^2 = 0$ \eqref{linear} can be an exact vacuum solution of the EoM for the nonlocal gravitational theory \eqref{action} or (\ref{tildeF}) for a specific but large class of nonlocal form factors. 

%%%
Specifically, the existence of G\"{o}del-type exact solutions depends on the choice of the entire function $\mathrm{H}_2$. 
According to the power counting super-renormalizability of the theory at quantum level, we focused on:
 \be\label{eq.H2summary}
{\rm H}_2(x)\equiv \alpha \int_0^{p(x)}\dd{z}\frac{1- e^{-z^n}}{z}\, , 
\ee
(see the subsection~\ref{sec.Hx} for more details about ${\rm H}_2$). 
Therefore, the existence of non-trivial solutions of the EoM can be rephrased as:\footnote{For the case $n = 1$ (see Subsubsection~\ref{sec.n=1}), an additional condition is required, namely that $p(x)$ must have at least a double root at $x = 0$.}:
\begin{mdframed}
	\centering 
	$n>1$ is odd, $\exists\, \zeta$ such that $p'(\zeta)=0$, $e^{{\rm H}(\zeta)}=1-\frac{9}{8}\lambda$ $\quad \Longleftrightarrow$  \quad G\"{o}del-type solution with $m^2=0\,, \,\, \omega^2=\frac{\zeta}{6}{\Lambda_*}^2$, 
\end{mdframed}
where $\lambda\equiv 1/\qty(2\kappa^2{\Lambda_*}^2)$ is a dimensionless constant related to the invariant fundamental mass scale ${\Lambda_*}$ of the theory and the gravitational constant $\kappa$.
{ We provided, at the end of Subsubsection~\ref{sec.n=1}, an explicit example of a form factor that admits the G\"odel-type solution with CTCs. In that example, we take $n = 1$ and choose $p(x) = x^2 \qty(2x - 3\zeta)$ with $\zeta > 0$.}

We conclude that while the super-normalizability excludes acausal G\"{o}del-type solutions in the presence of matter and negative cosmological constant, the very same quantum property is not enough to avoid the problem of CTCs in an empty Universe. However, in this respect, it is the matter itself to make the Universe causal by its own existence.

\appendix
\section{Nonlocal gravity in the Riemann-Ricci basis}
\label{RB}

In this section we derive the relation between  the theory n the Ricci-Weyl basis (\ref{action}) to the theory in the Ricci-Riemann basis (\ref{tildeF}). 
We start by recalling the following definition of the Weyl tensor in dimension $D$,
\be
C_{abcd}=R_{abcd}-\frac{2}{D-2}\qty(g_{a[c}R_{d]b}-g_{b[c}R_{d]a})+\frac{2}{(D-1)(D-2)}Rg_{a[c}g_{d]b} \, ,  
\ee
(notice that $C_{abcd}$ is traceless). Afterwards, we evaluate the Weyl square scalar, namely 
\be 
C_{abcd}C^{abcd}=R_{abcd}C^{abcd}=C_{abcd}R^{abcd}=\frac{2}{(D-1)(D-2)}R^2-\frac{4}{D-2}R_{ab}R^{ab}+R_{abcd}R^{abcd} \, .
\ee
Since $\nabla_cg_{ab}=0$ and $C_{abcd}$ is a linear function of the Riemann tensor $R_{abcd}$, we have 
%\begin{equation}
	%\begin{aligned}
	\be
	&&	\hspace{-0.35cm}
	\mathcal{L}_g = \frac{R-2\Lambda_{\rm{cc}}}{2\kappa^2} +R\gamma_{0}(\square)R+R_{ab}\gamma_{2}(\square)R^{ab}+C_{abcd}\gamma_{4}(\square)C^{abcd} \nonumber \\
		&& = \frac{R-2\Lambda_{\rm{cc}}}{2\kappa^2}   +R\gamma_{0}(\square)R+R_{ab}\gamma_{2}(\square)R^{ab}+\frac{2}{(D-1)(D-2)}R\gamma_{4}(\square)R-\frac{4}{D-2}R_{ab}\gamma_{4}(\square)R^{ab}+R_{abcd}\gamma_{4}(\square)R^{abcd} \nonumber \\
		&&
		= \frac{R-2\Lambda_{\rm{cc}}}{2\kappa^2} +R\qty[\gamma_{0}(\square)+\frac{2}{(D-1)(D-2)}\gamma_{4}(\square)]R+R_{ab}\qty[\gamma_{2}(\square)-\frac{4}{D-2}\gamma_{4}(\square)]R^{ab}+R_{abcd}\gamma_{4}(\square)R^{abcd} \, , 
		\label{redef}
		\ee
		which has to be equal to (\ref{tildeF}), namely 
		\be
		\frac{1}{2\kappa^2} \left( R-2\Lambda_{\rm{cc}} \right) +R\tilde\gamma_{0}(\square)R+R_{ab}\tilde\gamma_{2}(\square)R^{ab}+R_{abcd}\tilde\gamma_{4}(\square)R^{abcd} \, .
		\label{tildeF2}
	\ee
%so, we have 
Comparing the last step in (\ref{redef}) with the Lagrangian (\ref{tildeF2}), 
\be
\tilde\gamma_0=\gamma_0+\frac{2}{(D-1)(D-2)}\gamma_4\, , \quad \tilde\gamma_2=\gamma_2-\frac{4}{D-2}\gamma_4 \, , \qand \tilde\gamma_4=\gamma_4 \, . 
\ee
For $D=4$, we have the following relation between form factors $\gamma_i$ and $\tilde{\gamma}_i$, 
\begin{equation}
	\label{rienbasis}
	\gamma_2+2\gamma_4=(\gamma_2-2\gamma_4)+4\gamma_4=\tilde\gamma_2+4\tilde\gamma_4  \, .
\end{equation}

\section{Equations of motion of the theory in Ricci - Weyl basis %(\ref{action})
\label{RB2}
}
In this section we consider the following action for NLG, 
\begin{eqnarray}
\label{actmodtotRW}
S = \int \dd[4]x\sqrt{-g}\left[ \frac{1}{2\kappa^2}\left( 
R-2\Lambda_{\rm{cc}} \right) +R\gamma_{0}(\square)R+R_{ab}\gamma_{2}(\square)R^{ab}+C_{abcd}\gamma_{4}(\square)C^{abcd} \right] 
%\mathcal{L}_0
%
+S_{m}[g_{ab},\Psi] \,.
\label{actotRW}
%\label{L0}
\end{eqnarray}  
Varying the action \eqref{actmodtotRW} with respect to the metric, one can obtain the following field equations:
\begin{eqnarray}
\nonumber 
&& 
\hspace{-1cm}
E^{ab}=G^{ab}+\Lambda_{\mbox{cc}} g^{ab}+P_{1}^{ab}+P_{2}^{ab}+P_{3}^{ab}-2\Omega_{1}^{ab}+g^{ab}(g_{cd}\Omega^{cd}_{1}+\bar{\Omega}_{1})-2\Omega^{ab}_{2}\\
&& + g^{ab}(g_{cd}\Omega^{cd}_{2}+\bar{\Omega}_{2})-4\Delta^{ab}_{2}-2\Omega^{ab}_{3}+g^{ab}(g_{cd}\Omega^{cd}_{3}+\bar{\Omega}_{3})-8\Delta^{ab}_{3} 
- \kappa^{2}T^{ab} = 0 \, ,
\label{ac1}
\end{eqnarray} 
where the tensors $P_{i}^{ab}$ in (\ref{ac1}) are defined as follows, 
 \begin{eqnarray}
\nonumber
 P_{1}^{ab}&=&\kappa^2 \left[\left(4G^{ab}+g^{ab}R-4(\nabla^{a}\nabla^{b}-g^{ab}\square)\right)\gamma_0(\square)R\right] \, , \\
\nonumber P_{2}^{ab}&=&\kappa^2\bigg[4R^{{(}a}_{\,d}\gamma_2(\square)R^{{|}d{|}b{)}}-g^{ab}R^{cd}\gamma_2(\square)R_{cd}-4\nabla_{d}\nabla^{{(}b}(\gamma_2(\square)R^{{|}d{|}a{)}})+ 2\square(\gamma_2(\square)R^{ab})+2g^{ab}\nabla_{c}\nabla_{d}(\gamma_2(\square)R^{cd})\bigg] \, , \\
%\nonumber
 P_{3}^{ab}&=&\kappa^2\bigg[-g^{ab}C^{cdef}\gamma_4(\square)C_{cdef}+4C^{{(}a}_{\,\,cde}\gamma_4(\square)C^{b{)}cde} -4(R_{cd}+2\nabla_{c}\nabla_{d})(\gamma_4(\square)C^{{(}b{|}cd{|}a{)}})\bigg],
\label{EoM1}
\end{eqnarray}
while the tensors $\Omega^{ab}_{i}$ and $\tilde\Omega^{ab}_{i}$ in (\ref{ac1}) read:
\begin{eqnarray}	
\nonumber 
\Omega^{ab}_{1}&=&\kappa^2\sum_{n=1}^{\infty}\gamma_{0,n}{\frac{1}{\Lambda_*^{2n}}}\sum_{l=0}^{n-1}\nabla^{a}R^{(l)}\nabla^{b}R^{(n-l-1)}, \quad \bar{\Omega}_{1}=\kappa^2\sum_{n=1}^{\infty}\gamma_{0,n}{\frac{1}{\Lambda_*^{2n}}}\sum_{l=0}^{n-1}R^{(l)}R^{(n-l)} \, , \\
\nonumber\Omega^{ab}_{2}&=&\kappa^2\sum_{n=1}^{\infty}\gamma_{2,n}{\frac{1}{\Lambda_*^{2n}}}\sum_{l=0}^{n-1}(\nabla^{a}R^{cd(l)})(\nabla^{b}R_{cd}^{(n-l-1)}),\quad \bar{\Omega}_{2}=\kappa^2\sum_{n=1}^{\infty}\gamma_{2,n}{\frac{1}{\Lambda_*^{2n}}}\sum_{l=0}^{n-1}R^{cd(l)}R_{cd}^{(n-l)} \, , \\
\nonumber\Omega^{ab}_{3}&=&\kappa^2\sum_{n=1}^{\infty}\gamma_{4,n}{\frac{1}{\Lambda_*^{2n}}}\sum_{l=0}^{n-1}(\nabla^{a}C^{c(l)}_{\,\,def})(\nabla^{b}C_{c}^{\,\,def(n-l-1)}) \, \\
 \tilde{\Omega}_{3}&=&\kappa^2\sum_{n=1}^{\infty}\gamma_{4,n}{\frac{1}{\Lambda_*^{2n}}}\sum_{l=0}^{n-1}C^{a(l)}_{\,bcd}C^{\,\,bcd(n-l)}_{a}.
\label{OmegaA}
\ee
Finally, $\Delta^{ab}_{i}$ are:
\be
\nonumber\Delta^{ab}_{2}&=&{\kappa^2}\sum_{n=1}^{\infty}\gamma_{2,n}{\frac{1}{\Lambda_*^{2n}}}\sum_{l=0}^{n-1}\nabla^{c}\left(R_{dc}^{(l)}\nabla^{(a}R^{b)d(n-l-1)}-(\nabla^{(a}R_{dc})R^{b)d(n-l-1)}\right) \, , \\
\Delta^{ab}_{3}&=&{\kappa^2}\sum_{n=1}^{\infty}\gamma_{4,n}{\frac{1}{\Lambda_*^{2n}}}\sum_{l=0}^{n-1}\nabla^{c}\left(C_{\,\,cef}^{d(l)}\nabla^{(a}C^{\,\,b)ef(n-l-1)}_{d}-(\nabla^{(a}C_{\,\,cef}^{|d(l)|})C^{b)ef(n-l-1)}_{d}\right) ,
\label{OmegaD}
\end{eqnarray}
here we are used the notation 
$A^{(l)}\equiv \square^{l}A$.

\section*{Acknowledgments}
This work was supported by the National Natural Science Foundation of China (NSFC), Grant No.~W2531002.

\end{document}